\documentclass[runningheads]{llncs}
\usepackage[T1]{fontenc}
\usepackage{xcolor}

\usepackage{xspace}
\newcommand{\flexisaga}{Flexi\-SAGA\xspace}

\usepackage{graphicx}
\usepackage{subfigure}
\usepackage{orcidlink}

\begin{document}
\title{FlexiSAGA: A Flexible Systolic Array GEMM Accelerator for Sparse and Dense Processing}
\titlerunning{FlexiSAGA}

%
\author{Mika Markus Müller\textsuperscript{*}\orcidlink{0009-0003-3471-7560} \and
Konstantin Lübeck\textsuperscript{*}\orcidlink{0000-0002-2701-5881}\and \\
Alexander Louis-Ferdinand Jung\orcidlink{0000-0001-5702-5768}\and
Jannik Steinmetz\orcidlink{0009-0003-7193-5620}\and \\
Oliver Bringmann\orcidlink{0000-0002-1615-507X}
}
\authorrunning{M. M. Müller and K. Lübeck et al.}
%
\institute{Embedded Systems, University of Tübingen, Tübingen, Germany\\\quad\\
\textsuperscript{*}These authors contributed equally to this work.
}
\maketitle              
\vspace{-5mm}
\begin{abstract}
Artificial Intelligence (AI) algorithms, such as Deep Neural Networks (DNNs), have become an important tool for a wide range of applications, from computer vision to natural language processing. However, the computational complexity of DNN inference poses a significant challenge, particularly for processing on resource-constrained edge devices. One promising approach to address this challenge is the exploitation of sparsity in DNN operator weights.

In this work, we present FlexiSAGA, an architecturally configurable and dataflow-flexible AI hardware accelerator for the sparse and dense processing of general matrix multiplications (GEMMs). FlexiSAGA supports seven different sparse and dense dataflows, enabling efficient processing of resource intensive DNN operators. Additionally, we propose a DNN pruning method specifically tailored towards the FlexiSAGA architecture, allowing for near-optimal processing of dense and sparse convolution and fully-connected operators, facilitating a DNN/HW co-design flow. Our results show a whole DNN sparse-over-dense inference speedup ranging from 1.41 up to 4.28, outperforming commercial and literature-reported accelerator platforms.
\end{abstract}

\section{Introduction}
In recent years, the deployment of AI workloads, such as Deep Neural Networks (DNNs), has shifted away from datacenters to resource-constrained edge devices due to privacy concerns, real-time requirements and costs. To fulfil these non-functional requirements, specialized AI accelerators are often necessary, as they can process data locally and faster than conventional microcontrollers while maintaining a small area and energy footprint.

When using an AI accelerator, the goal is often to efficiently process the most computationally intensive DNN operators, like convolution (CONV) and fully-connected (FC). While the FC operator implements a general matrix multiplication (GEMM), the CONV operator can be converted into a GEMM by applying an im2col transformation \cite{im2col2005}. This enables the usage of the same architecture and mapping approach to process both operators. The GEMM can be further optimized by splitting it into smaller tiles, which allows for improved cache utilization, increased parallelism, and memory bandwidth optimization.

State-of-the-art GEMM accelerators \cite{eyeriss2017,ultratrail2020,gemmini2021} use systolic arrays (SAs) to efficiently compute tiled GEMMs as they do not have the drawbacks of Von Neumann architectures \cite{nonvonneumannsurvey2019}. SAs enable parallel processing where some data is streamed, while the other data is kept stationary to reach maximum reuse of the stationary data which minimizes costly memory accesses. The dataflow determines which type of data remains stationary. There are three common dataflows used in SAs: output stationary (OS), weight stationary (WS) and input stationary (IS). For different matrix and tile sizes used for the tiled GEMM these dataflows can reach vastly different runtimes depending on the architecture.

To further improve the processing and memory footprint of DNNs, pruning has become a popular technique \cite{pruningsurvey2024}. Pruning involves selectively removing less important DNN weights by replacing them with zeros, and thereby reducing the overall model complexity without significantly impacting its accuracy, which results in sparse weight matrices and therefore in sparse GEMMs. Several AI accelerators have been proposed which employ sparsity-centric optimizations to efficiently process sparse GEMMs \cite{eyeriss2017,scnn2017,sparten2019,spots2022}. However, a common limitation of many AI accelerators is that they tend to only process the sparse GEMMs using a single dataflow, which may not be the most efficient one for all DNN operators.

In this paper, we present \flexisaga, a flexible and configurable systolic array for GEMM acceleration, which can process both dense and sparse GEMMs using seven different dataflows to achieve a higher flexibility and higher sparse-over-dense speedups than many other accelerators when processing AI workloads. To reach this goal, sparsity is exploited solely within the weight matrix. This can be done at the time of deployment, i.e. without affecting the on-device DNN processing time, compared to the dynamic identification of zeros within the input matrices at runtime. We use the two-stage bitmap format \cite{spots2022} and we present a custom sparse format to efficiently store and decode weight matrix tiles. Furthermore, we introduce a structured pruning technique based on \cite{ssl2016}. The custom sparse format and the pruning technique are tailored specifically to the \flexisaga architecture, enabling a DNN/HW co-design flow.

We evaluate FlexiSAGA with a number of representative DNNs (AlexNet \cite{alexnet2017}, VGG16 \cite{vgg2015}, GoogLeNet \cite{googlenet2015}, and ResNet50 \cite{resnet2015}) pruned and deployed on differently sized FlexiSAGA instances and compare the achieved sparse-over-dense speedups to an Intel Xeon CPU, the Nvidia Orin ARM CPU and GPU, and the SCNN \cite{scnn2017} and SparTen \cite{sparten2019} accelerators. These results show a whole DNN sparse-over-dense speedup of 1.41 up to 4.28. Additionally, we conduct a design space exploration (DSE) to find the near-optimal combination of dataflow, pruning and architectural parameters to minimize the DNN operator runtimes.

\section{Related Work}
Several AI accelerators and pruning frameworks for processing sparse matrices have been proposed. The SPOTS accelerator \cite{spots2022} has an integrated im2col unit which transforms CONV input and weight tensors on-the-fly to process them as tiled GEMM. Additionally, the authors introduce a two-stage bitmap format for storing sparse matrices pruned using structured sparsity learning \cite{ssl2016}. Their accelerator shows a speedup of up to 20 for whole DNNs compared to the dense execution on a CPU and a speedup of up to 1.86 compared to other accelerator architectures presented in \cite{eyeriss2017,gemmini2021}, supporting dense and sparse processing, using a reconfigurable systolic array composed of 512 processing elements.

SCNN \cite{scnn2017} focuses on the processing of CONV operators. Exploiting input and weight sparsity, the authors present a whole DNN speedup of up to 3.52 for VGG16 compared to the dense processing on the same architecture utilizing 64 processing elements and using a novel input stationary dataflow.

SparTen \cite{sparten2019} uses bitmap encoding to store sparse input and weight matrices. The results show a sparse-over-dense speedup of up to 15.5 for CONV operators using up to 64 compute clusters. 

DeepSparse \cite{deepsparse2023} is a ``Sparsity-aware deep learning inference runtime for CPUs''\footnote{\texttt{https://github.com/neuralmagic/deepsparse} (accessed March 13, 2025).} which provides an automatic deployment for pruned PyTorch \cite{pytorch2019} models. The authors report a whole DNN sparse-over-dense speedup of up to 8.2 for 32bit floating-point models running on an Intel Xeon CPU.

Nvidia's 2:4 sparsity \cite{nvidiaasp2021} is specifically tailored towards the Nvidia Ampere microarchitecture. 2:4 sparsity pruning splits the weight matrix into row vectors containing four elements. In each vector, the two smallest values are set to zero. After pruning, the DNNs are fine-tuned. They report a sparse-over-dense speedup of up to 2 for an Nvidia A100 GPU using 16bit floating-point models.

We compare our FlexiSAGA architecture together with the proposed DNN pruning method to the CONV operator sparse-over-dense speedup of the one-sided SCNN and SparTen accelerators presented in \cite{sparten2019} as they provide detailed per DNN operator results. Additionally, we compare our whole DNN sparse-over-dense speedup to DeepSparse and TensorRT, run on an Intel Xeon CPU, Nvidia Orin ARM CPU and GPU respectively.

\section{Sparse Matrix Formats}
\begin{figure}[!t]
    \centering
    \subfigure[]{
        \includegraphics[width=0.4\textwidth]{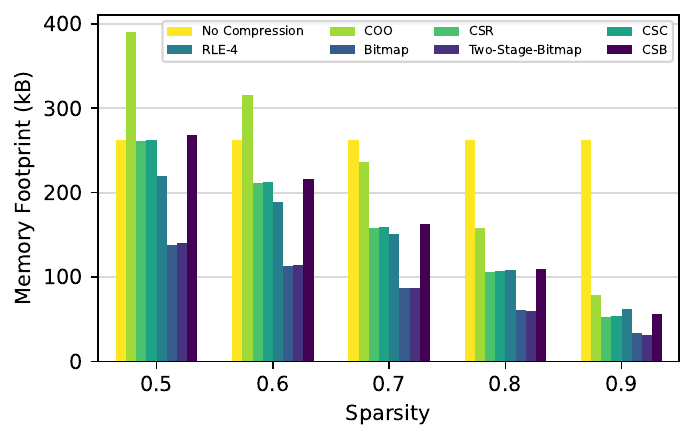}
    }
    \subfigure[]{
        \includegraphics[width=0.16\textwidth]{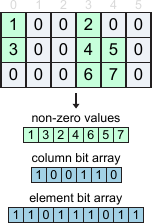}
    }
    \subfigure[]{
        \includegraphics[width=0.16\textwidth]{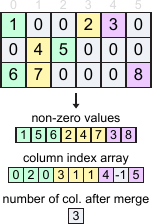}
    }
    \caption{\textbf{(a)} Memory footprint comparison of different sparse matrix formats and no compression for a 128$\times$512 matrix of 32bit values with varying sparsities and uniformly distributed zero elements. \textbf{(b)} Two-stage bitmap format example. \textbf{(c)} Compressed sparse block (CSB) format example.}
    \label{fig:sparse_formats}
\end{figure}

An important factor for processing sparse GEMM operations is in which format sparse matrices are stored. A sparse matrix format should have a good compression ratio to minimize the memory footprint and allow for efficient decompression such that it does not slow down the processing of a GEMM operation. In the following section, different sparse matrix formats are briefly introduced and compared to each other. Additionally, we present the compressed sparse block (CSB) format, which is tailored towards the sparse GEMM processing of \flexisaga. Fig.~\ref{fig:sparse_formats}(a) shows a comparison of the different sparse matrix formats for a 128$\times$512 matrix with varying sparsities and uniformly distributed zeros.

The compressed sparse row (CSR) format stores the non-zero elements of a matrix in an array, along with two additional arrays that store the column indices and the row pointers. The row pointers array indicates the starting index of each row in the non-zero elements array. Similarly to CSR is the compressed sparse column (CSC) format, but it stores a matrix by columns instead of rows. The non-zero elements are also stored in an array, along with the row indices and column pointers stored in two additional arrays. The coordinate format (COO) stores the non-zero elements of a matrix as a list of (row, column, value) tuples, which is easy to implement but comes with significant storage overhead. The Run-Length Encoded 4bit (RLE-4) format  represents a matrix as a sequence of 4bit codes, where each code indicates the length of a run of zeros followed by a non-zero element. This format can be very compact for matrices with long runs of zeros, but it may be less efficient for matrices with more uniform sparse distributions. 
The bitmap format also contains an array for the non-zero elements, but additionally defines a bit array, where for every element is stored whether it is non-zero or zero indicated by a 1 and 0 respectively.

The two-stage bitmap format introduced in \cite{spots2022} uses an array containing all non-zero elements of a matrix and two bit arrays which encode the coordinates of non-zero elements. The first array, called the column bit array, contains as many bits as columns in a matrix. A 0 in the column bit array encodes that the corresponding matrix column contains only zero elements, a 1 encodes that the corresponding column contains non-zero elements. The second bit array, called the element bit array, encodes for each non-zero column which elements are non-zero. This allows for a high compression ratio and efficient decompression, where loading of entire zero columns can be skipped. Fig.~\ref{fig:sparse_formats}(b) shows an example encoding for a 3$\times$6 sparse matrix.

The two-stage bitmap format demonstrates a good compression ratio, even for low sparsities (see Fig.~\ref{fig:sparse_formats}(a)). However, the occurrence of entire zero columns of height $n$ with uniformly distributed zero elements and sparsity $s$ is described by the Bernoulli process ${n\choose n}s^n(1-s)^{n-n} = s^n$ and is consequently very low for large $n$. Therefore, we introduce the compressed sparse block (CSB) format, which allows merging multiple non-zero columns with high sparsity into one. Similarly to the two-stage bitmap format, the CSB format uses one array to store all non-zero elements of a matrix, additionally for each non-zero element the column index is stored. The row index is implicitly encoded in the order of the column index array.
For each column starting from the first, we use greedy search to find matching columns to merge with. Columns match if the position of the non-zero elements of one column matches the position of the zero elements in the other column, while it is allowed that zero elements of one column can match zero elements of the other column. Columns only containing zeros are not merged and are skipped entirely. The resulting number of columns after the merge is stored together with the non-zero elements and the column index array. Fig.~\ref{fig:sparse_formats}(c) shows an example of the CSB format. Because of the column index array, the memory footprint for the CSB format is higher than for the bitmap formats. However, it allows for skipping more than just zero columns because combined columns are loaded as a single column.

FlexiSAGA utilizes the two-stage bitmap format and the CSB format to efficiently store and process sparse matrices, which is described in detail in the following section.

\section{FlexiSAGA Architecture}
\begin{figure}[!t]
	\centering
    \subfigure[]{
        \includegraphics[width=0.42\textwidth]{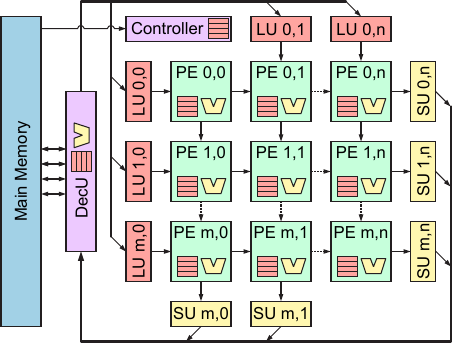}
    }
    \subfigure[]{
        \includegraphics[width=0.4\textwidth]{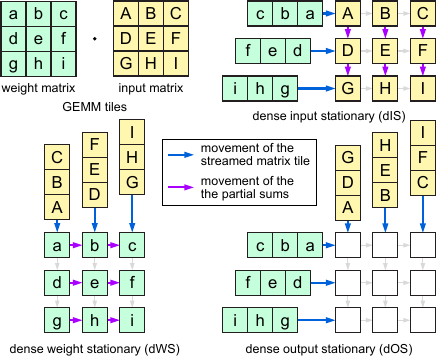}
    }
	\caption{\textbf{(a)} Block diagram of the FlexiSAGA architecture. \textbf{(b)} Visualization of the three dense tiled GEMM dataflows supported by the FlexiSAGA architecture.}
    \label{fig:flexisaga_architecture_and_dense_dataflows}
\end{figure}

Fig.~\ref{fig:flexisaga_architecture_and_dense_dataflows} shows the block diagram of the proposed \flexisaga architecture. \flexisaga is designed as a systolic array (SA) whose main components are Processing Elements (PEs) organized in a 2D grid, configurable in its height and width. Each PE has a register file containing nine registers of configurable data word width to store stationary data and partial sums. An arithmetic logic unit (ALU) supporting floating-point and integer operations inside each PE can access the register file to execute move, multiply, addition, and multiply-accumulate operations. Furthermore, the ALU of each PE can also write into the register files of its neighboring PEs to its right and below, which facilitates data transfer through the SA. To load data from the main memory into the SA, each PE in the left column and in the top row is connected to a load unit (LU) which performs memory read transactions and writes the read data words into the register file of the connected PE. All LUs are connected to the decompression unit (DecU) which acts as an arbiter for main memory accesses. When a dense GEMM is processed, the DecU forwards all read transactions to the main memory. In case of a sparse GEMM the DecU checks if the accessed value is declared as a zero in the sparse representation of the weight matrix tile and emits a zero instead of forwarding the read transaction to the main memory. The data width and the number of ports connecting the main memory to the DecU is configurable. The PEs on the bottom and the right column are connected to store units (SUs) which read data from the PE register files and write them into the memory, again through the DecU. The behavior of all components is globally set by a programmable controller, which contains a schedule for the dense or sparse GEMM. In case of a sparse GEMM the controller is programmed to skip the processing of entire zero columns or rows of the weight matrix. The following sections detail how dense and sparse GEMMs are processed by the \flexisaga architecture.

\subsection{Dense Tiled GEMM Processing}

\flexisaga supports the three common dense GEMM dataflows: input stationary (dIS), weight stationary (dWS), and output stationary (dOS). Before processing a dense GEMM on the FlexiSAGA architecture, the weight and input matrices are split into tiles of the size of the SA. Fig.~\ref{fig:flexisaga_architecture_and_dense_dataflows}(b) presents the different dense tiled GEMM dataflows. When using the dIS dataflow, the elements of the weight matrix tile are streamed into the SA, while the elements of the input matrix tile are held stationary in the PEs. Each PE performs a multiply-accumulate operation between weight and input elements, accumulating the partial sums, which are then propagated through the array to compute the output tile. The dWS dataflow is similar to the dIS dataflow, but the roles of weight and input matrix tiles are reversed. In the dOS dataflow, the elements of the weight and input tiles are streamed into the SA, and each PE performs a multiply-accumulate operation, accumulating the partial sums in a local register. The output tile is then read from the PE registers after all the partial sums have been accumulated.

\subsection{Sparse Tiled GEMM Processing}

In addition to the dense tiled GEMM processing, FlexiSAGA supports output stationary (sOS), weight stationary (sWS), and input stationary (sIS) dataflows for sparse tiled GEMMs. All three sparse dataflows use the two-stage bitmap format to compress the weight matrix tiles. Additionally, for the output stationary dataflow we propose a variant using the CSB format to compress the weight matrix tiles (csOS). The following sections detail all four sparse dataflows. The tiles of the weight matrices are stored in a compressed format in the main memory before the inference of a DNN. The input matrices are stored in row-major order in the main memory, and the addresses for accessing the corresponding input tile elements for a weight tile are generated by the \flexisaga controller. The elements of the output matrix are stored in row-major order in the main memory and serve as input matrix for the succeeding DNN operator.

\begin{figure}[!t]
	\centering
	\includegraphics[width=0.8\textwidth]{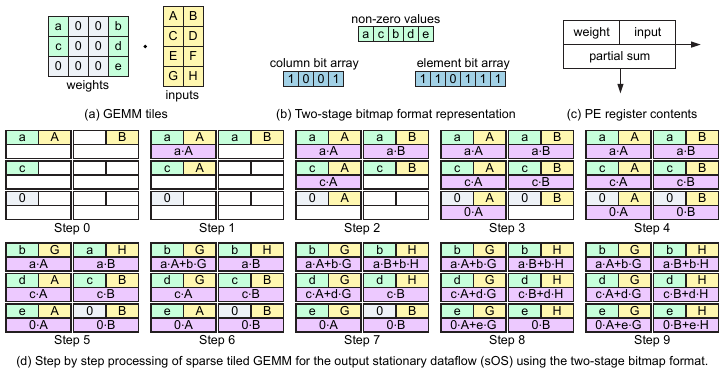}
	\caption{Example for processing single tile using the sparse tiled GEMM output stationary dataflow (sOS) on a FlexiSAGA architecture of size 2$\times$3.}
    \label{fig:sos_dataflow}
\end{figure}
\subsubsection{Sparse Output Stationary Dataflow (sOS)}
Fig.~\ref{fig:sos_dataflow} presents an example of processing of a single GEMM tile using the sOS dataflow. All weight matrix tiles are compressed using the two-stage bitmap format (see Fig.~\ref{fig:sos_dataflow}(b)). While the weight matrix tile consists of 12 elements in its uncompressed form, storing it in the two-stage bitmap format allows us to only read seven data words to access the whole tile. Fig.~\ref{fig:sos_dataflow}(d) shows the step by step processing for the sOS dataflow. Each step shows the register contents of all six PEs, while Fig.~\ref{fig:sos_dataflow}(c) provides a legend for the PE register contents. In step 0, the first column of the weight tile (\texttt{a,c,0}) is loaded into the registers of the PEs in the left column and the first row of the input tile (\texttt{A,B}) is loaded into PEs in the top row. In step 1, the top left PE calculates the first partial sum and stores it in its register, additionally this PE copies the input element \texttt{A} into the PE below it and the weight element \texttt{a} into the PE to the right of it. In step 2, the top right PE and the middle left PE calculate their partial sums and copy the input and weight elements to their neighbor PEs below and to the right (if they exist). This is repeated until step 4, in which all PEs contain a partial sum and each element of the first weight tile column has been multiplied with each element of the first input tile row. In steps 5 to 9, the last column of the weight tile (\texttt{b,d,e}) and the last row of the input tile (\texttt{G,H}) are loaded into the PEs and the partial sums of the multiplications are accumulated. The PE registers now contain all non-zero elements of the output tile, which are then forwarded to the main memory.

\begin{figure}[!t]
	\centering
	\includegraphics[width=0.8\textwidth]{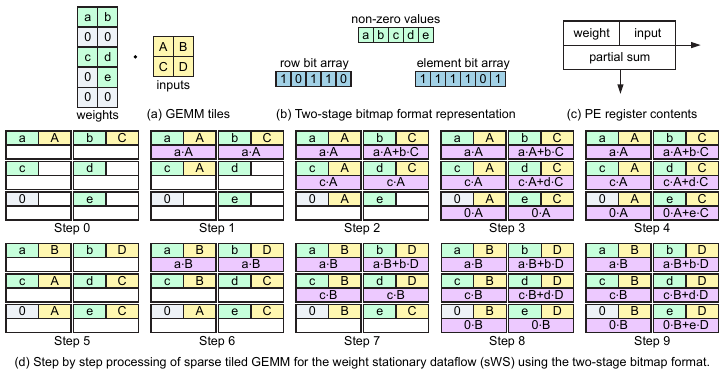}
	\caption{Example for processing single tile using the sparse tiled GEMM weight stationary dataflow (sWS) on a FlexiSAGA architecture of size 2$\times$3.}
    \label{fig:sws_dataflow}
\end{figure}
\subsubsection{Sparse Weight Stationary Dataflow (sWS)}
Fig.~\ref{fig:sws_dataflow} presents an example of processing of a single GEMM tile using the sWS dataflow. Fig.~\ref{fig:sws_dataflow}(d) shows the PE register contents for each step of the sWS dataflow processing. In step 0 the whole weight tile is loaded into the SA and stays there until all elements of the output tile have been calculated. In steps 1 to 4, the first column of the input tile is propagated vertically through the SA and the first column of the output tile is accumulated in the right PE column and then forwarded to the main memory. In steps 5 to 9, the partial sums are cleared and the second column of the input tile is propagated through the SA in the same manner as the first column. Similarly to the sOS dataflow in this example only seven data words need to be read from the main memory to process a weight tile containing ten elements.

\begin{figure}[!t]
	\centering
	\includegraphics[width=0.8\textwidth]{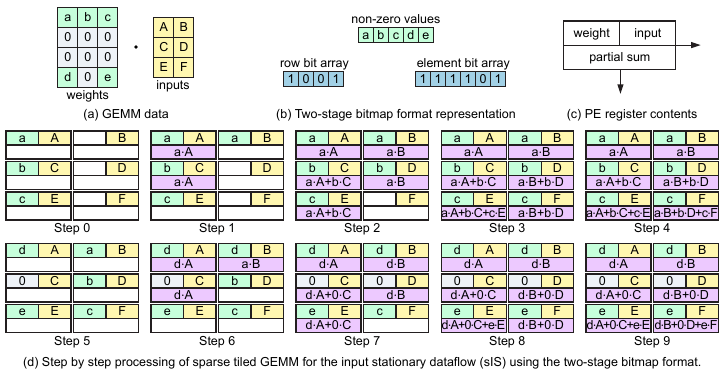}
	\caption{Example for processing single tile using the sparse tiled GEMM input stationary dataflow (sIS) on a FlexiSAGA architecture of size 2$\times$3.}
    \label{fig:sis_dataflow}
\end{figure}
\subsubsection{Sparse Input Stationary Dataflow (sIS)}
Fig.~\ref{fig:sis_dataflow} presents an example of processing of a single GEMM tile using the sIS dataflow. Fig.~\ref{fig:sis_dataflow}(d) shows the PE register contents for each step of the sIS dataflow processing. In step 0 the whole input tile is loaded into the SA and stays there until all output tile elements have been calculated. 
Additionally, the first row of the weight tile is loaded into the registers of the left PE column. 
In steps 1 to 4, the weight tile row is horizontally propagated through the SA and the first output tile row is accumulated in the bottom PE row and then forwarded to the main memory. In steps 5 to 9, the partial sums are cleared and the next non-zero row of the weight tile is propagated through the SA in the same manner as the first row. As seen before, the two-stage bitmap format drastically reduces the amount of data words which have to be read from the main memory for the weight tile.

\begin{figure}[!t]
	\centering
	\includegraphics[width=0.8\textwidth]{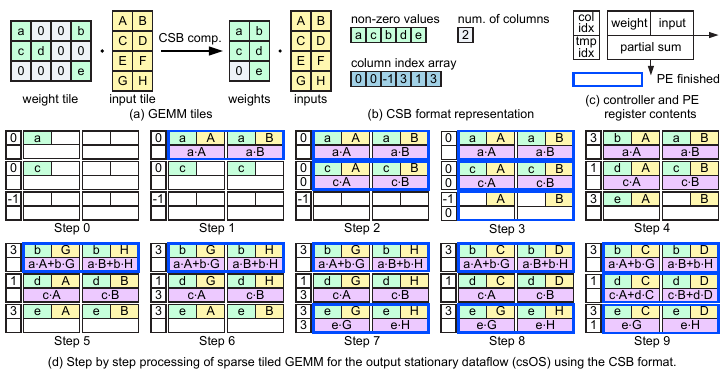}
	\caption{Example for processing single tile using the sparse tiled GEMM output stationary dataflow (csOS) using the CSB format on a FlexiSAGA architecture of size 2$\times$3.}
    \label{fig:csos_dataflow}
\end{figure}

\subsubsection{Sparse Output Stationary Dataflow (csOS) using the CSB format}
In contrast to the sOS, sWS, and sIS dataflows which utilize the two-stage bitmap format to compress the weight tiles, the csOS dataflow uses the CSB format. Since the CSB format merges multiple weight tile columns, the input tile rows and weight tile columns cannot simply be propagated through the SA as input and weight elements might not match anymore. Therefore, the \flexisaga controller tracks the column index of the weights the PE row currently stores to identify if a multiply-accumulate operation is needed.

Fig.~\ref{fig:csos_dataflow}(d) shows the step by step processing for the csOS dataflow. Each step shows the register contents of all six PEs together with the column index and a temporary index for each PE row stored in the controller, while Fig.~\ref{fig:csos_dataflow}(c) provides a legend for the controller and PE register contents. In step 0, the first column of the weight tile is loaded into the registers of the left PEs while the column index for each PE row is stored in the controller. In step 1 the first row of the input tile is loaded into the first PE row because the column index for the first PE row is set to 0, which corresponds to the first row of the input tile. Additionally, weights of the left PE column are copied to the right PE column and PEs in the first row calculate their partial sums. Because the PEs in the first row have finished their processing for the current weights, the PEs are marked with a blue outline. In step 2, the input elements are propagated vertically to the next PE row and the column index of the first row is copied to the temporary index of the second row and the PEs in the second row compute their partial sums. In step 3, the inputs are further propagated, however, the PEs in the last row do not have to execute a multiply-accumulate operation because the column index is set to -1 which indicates a zero weight. Now all rows are marked as finished, which implies that in step 4 the next weight column is loaded into the left PE column. Step 5 is analogous to step 1. In step 6, inputs are propagated vertically from the first to the second PE row. Now the column index and the temporary index in the second PE row do not match, which means no computation has to be done. Step 7 is analogous to step 2. After step 7, the second PE row remains unmarked, which means the PEs contain weights for a different input row. In step 8, the inputs corresponding to the column index stored for the second PE row are loaded into the SA and propagated to the second PE row. In step 9 the PE register files now contain all non-zero elements of the output tile, which are then forwarded to the main memory. If the two-stage bitmap format had been applied to the weight tile presented in Fig.~\ref{fig:csos_dataflow}(a), an additional weight column would be loaded when using the sOS dataflow resulting in more data movement.

\section{Deep Neural Network Pruning}
\label{sec:pruning}
To exploit sparsity during the inference of DNNs, they have to be pruned before deployment. The simplest technique is global unstructured pruning, in which a global sparsity $s$ for a whole pre-trained DNN is set. Then a proportion $s$ of all weights $w_i$ which have the smallest $l^1$-norm $|w_i|$ are set to zero. Even though this technique can lead to runtime improvements, there is often a considerable accuracy loss and an imbalanced sparsity distribution across all weights. Additionally, the unstructured sparsity may lead to a complex control flow, degrading the runtime improvements. 

To alleviate those drawbacks, we implemented a pruning technique based on structured sparsity learning \cite{ssl2016} in PyTorch \cite{pytorch2019} which allows us to introduce zero rows and columns in the DNN weight matrix tiles tailored towards the two-stage bitmap and CSB format. Firstly, the DNNs are trained using a standard training loop until they reach the desired accuracy $a$. Secondly, the prunable DNN operators, in our case CONV and FC, are grouped by operator type. Each group $j$ gets assigned a sparsity $s_j$. 
Afterward, each weight tensor of the CONV group is transformed into a matrix using an im2col transformation and split into tiles of the desired size. The FC weights can be split into tiles directly.
Those tiles are further split into row or column vectors $\vec{w}_i \in \mathbf{W}_j$ of length $n$ corresponding to the tile width or height. For each group $j$, the proportion $s_j$ of $\vec{w}_i \in \mathbf{W}_j$ with the smallest $l^2$-norm $||\vec{w}_i||_2 = \sqrt{\sum^n_{k=0} w_k^2}$ are set to zero. Then the DNN is trained again until it reaches accuracy $a-\epsilon$ with $\epsilon \geq 0$ while the pruned vectors stay zero. When the accuracy $a-\epsilon$ is reached, $s_j$ is increased by $\delta_j$ and the weights in each group are pruned again. This is repeated until the DNN training cannot reach accuracy $a - \epsilon$ anymore after a fixed amount of training epochs.

\section{Results}
\subsection{Experimental Setup}
To evaluate the FlexiSAGA architecture, we implemented a configurable virtual prototype (VP) using the Amaranth HDL \cite{amaranth2025} which allows for translation into a register-transfer level (RTL) representation for the integrated RTL simulator or into synthesizable Verilog. For the runtime results presented in this work, we used the Amaranth HDL RTL simulator together with a cycle-approximate SRAM memory model with unit read and write latencies and eight ports spread across multiple banks and a port width of 32bit adopted from the UltraTrail \cite{ultratrail2020} memory architecture. The PE's register files are set to a width of 32bit, and the PE ALUs support 32bit floating-point operations.

\begin{figure}[!t]
	\centering
	\includegraphics[width=1.0\textwidth]{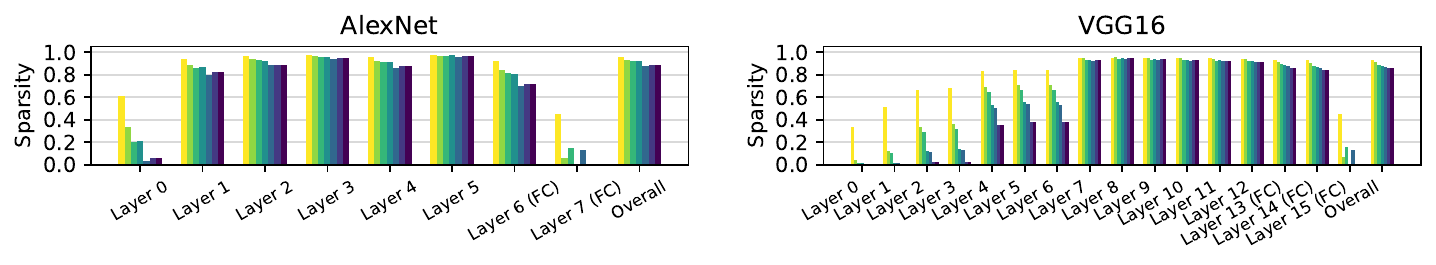}
	\includegraphics[width=1.0\textwidth]{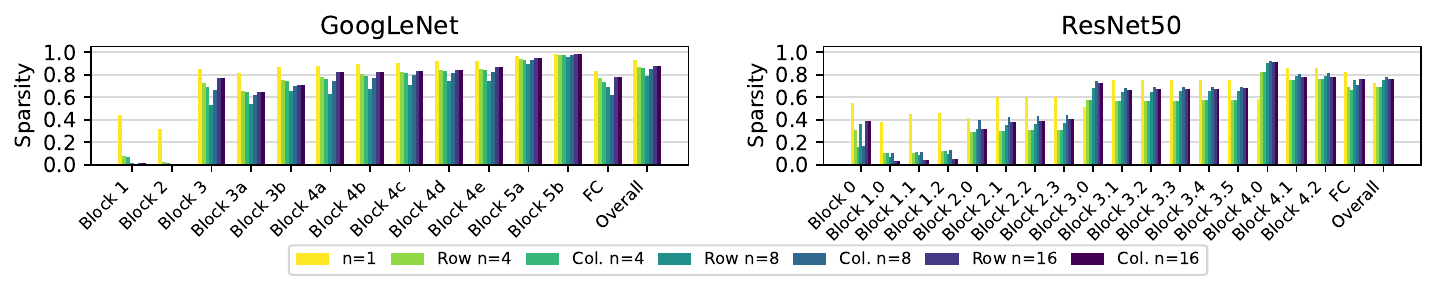}
	\caption{Operator and operator block sparsities for different DNNs and vector lengths $n$ and vector orientations.}
    \label{fig:dnn_sparsities}
\end{figure}

We use the PyTorch reference implementations of AlexNet \cite{alexnet2017}, VGG16 \cite{vgg2015}, ResNet50 \cite{resnet2015}, and GoogLeNet \cite{googlenet2015} and trained them for classification of the CIFAR-10 dataset with the following top-1 accuracies $a$: AlexNet 0.86, VGG16 0.89, ResNet50 0.89, and GoogLeNet 0.91 using 32bit floating-point weights. Those DNNs were chosen to compare against the results presented in \cite{sparten2019}. After training, the DNNs were pruned using the technique introduced in section \ref{sec:pruning} with an initial group sparsity $s_j = 0.7$, a sparsity progression $\delta_j = 0.01$ for all groups $j$, an $\epsilon = a\cdot0.02$, and a stochastic gradient descent optimizer which allows for a maximum top-1 accuracy drop of $2\%$ which is in accordance with the accuracies for pruned DNNs reported in \cite{spots2022}. Fig.~\ref{fig:dnn_sparsities} shows the operator and overall sparsities after pruning, for each DNN using different vector lengths $n$ (1, 4, 8, and 16) and row or column orientation. Since ResNet50 consists of 109 CONV and FC operators and GoogLeNet consists of 115 operators, their sparsity results are presented in operator blocks. Pruning with vector length $n=1$ shows the highest sparsities across all DNNs, however, $n=1$ does not introduce any structured sparsity. With an increasing $n$, the first couple of operators for all DNNs and the last FC operator of AlexNet and VGG16 drastically decrease in sparsity. The overall sparsity shows only slight variations for different vector lengths and orientations and remains above 0.75 for AlexNet, VGG16, and GoogLeNet which is comparable to the sparsities reported for the same DNNs in \cite{spots2022}.

To compare the sparse-over-dense inference speedup to other architectures, we used the DeepSparse \cite{deepsparse2022} inference engine and deployed the four DNNs unpruned (dense) and pruned (sparse) onto an Intel Xeon Platinum 8168 CPU with 24 cores and 48 threads and the Nvidia Jetson AGX Orin ARM Cortex-A78AE CPU with 12 cores. Each DNN was run using a single core and multiple cores/threads. The sparse variant of the DNNs were pruned with $\epsilon = a\cdot 0.02$ and $n = 1$, as it provides the highest sparsity and the highest speedup when using DeepSparse and limiting the top-1 accuracy drop to 2\%. Additionally, we used Nvidia's 2:4 sparsity \cite{nvidiaasp2021} to prune all four DNNs with a maximum top-1 accuracy drop of 2\% and deployed the dense and sparse 16bit floating-point variants onto the Nvidia Jetson AGX Orin Ampere GPU using TensorRT. For the DeepSparse and TensorRT runtime measurements, we collected multiple samples to calculate the mean to accommodate for interferences of the operating system. Moreover, we compared the operator-wise sparse-over-dense speedup to the SCNN and SparTen architectures using the results reported in \cite{sparten2019}. 

\subsection{Whole DNN Inference Evaluation}

\begin{figure}[!t]
	\centering
    \subfigure[]{
        \includegraphics[width=0.57\textwidth]{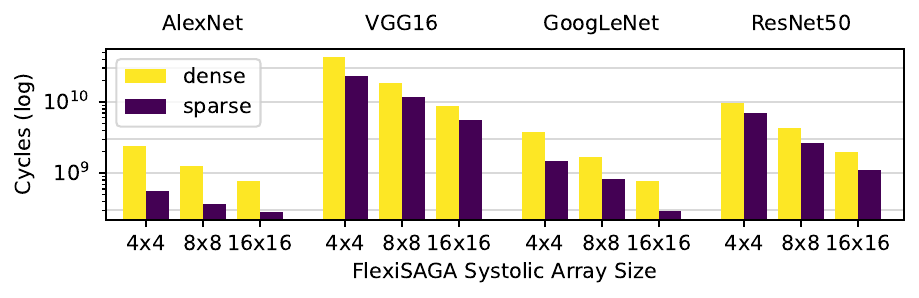}
    }
    \subfigure[]{
        \includegraphics[width=0.31\textwidth]{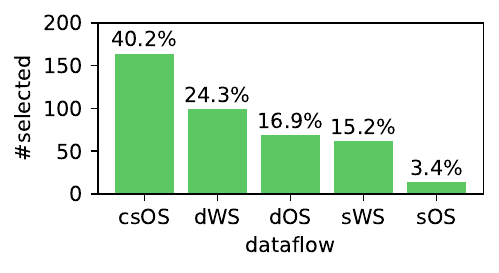}
    }
	\caption{\textbf{(a)} Comparison of whole DNN runtimes in clock cycles for different FlexiSAGA systolic array sizes. \textbf{(b)} Distribution of selected dataflows with minimal runtime per DNN operator across all DNNs and all FlexiSAGA sizes.}
    \label{fig:flexisaga_dnn_runtimes}
\end{figure}

Fig.~\ref{fig:flexisaga_dnn_runtimes}(a) presents the runtime in clock cycles for the four DNNs deployed onto the FlexiSAGA architecture of different sizes (4$\times$4, 8$\times$8, and 16$\times$16). Dense represents the clock cycle sum of the unpruned CONV and FC operators, while sparse represents the clock cycle sum of the pruned operators. The DNNs were pruned with vector length $n$ set to the systolic array size in row and column orientation. For each operator, the dataflow with the minimal runtime (dOS, dWS, dIS, sOS, sWS, sIS, or csOS) was chosen by measuring all different variants. For the pruned DNNs the vector orientation is the same for all operators. For all DNNs, the runtime decreases when quadrupling the SA size. The mean dense and sparse speedup across all DNNs when increasing the SA size is 2.1 and 2.07 respectively. Even though the number of PEs scales quadratically, the memory interface scales only linearly because only the outer PEs have access to the main memory. This linear scaling is reflected in the mean dense and sparse speedup.

\begin{figure}[!t]
	\centering
	\includegraphics[width=1.0\textwidth]{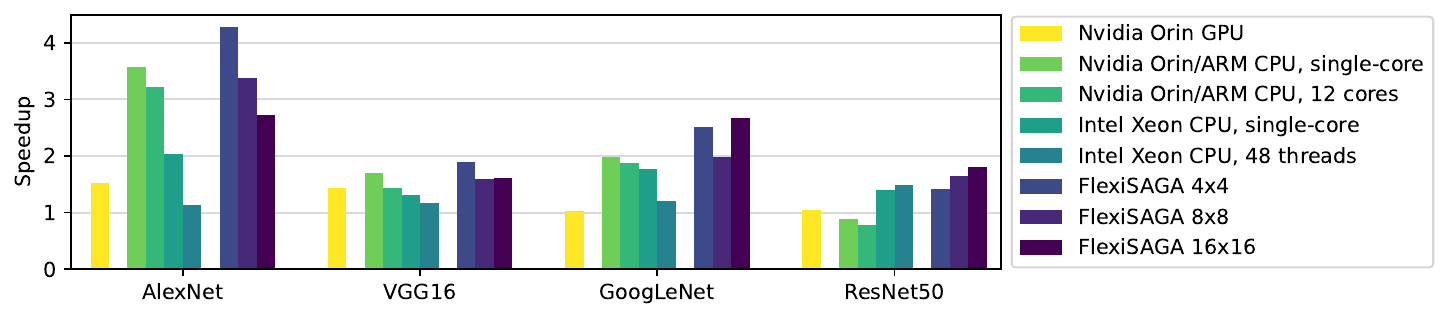}
	\caption{Whole DNN sparse-over-dense speedup comparison between an Nvidia Orin GPU (2:4 sparsity), Nvidia Orin ARM CPU and Intel Xeon CPU (DeepSparse), and FlexiSAGA with different systolic array sizes.}
    \label{fig:dnn_sparse_vs_dense_speedup_per_platform}
\end{figure}

Fig.~\ref{fig:dnn_sparse_vs_dense_speedup_per_platform} shows the sparse-over-dense speedup comparison for the Nvidia Orin GPU, Nvidia Orin ARM CPU, Intel Xeon CPU, and FlexiSAGA of different sizes. While FlexiSAGA only supports the CONV and FC operators, the other platforms support all operators of the deployed DNNs. However, the runtime of activation, pooling and element-wise operators is very small compared to the CONV and FC operators. AlexNet generally shows the best sparse-over-dense speedup for all platforms, this can be attributed to the high sparsity that can be achieved across all operators when pruning AlexNet (see Fig.~\ref{fig:dnn_sparsities}). ResNet50 shows the lowest speedup across all architectures because ResNet50 has the lowest overall sparsity and consists of 109 CONV and FC operators, which have smaller inputs and weights than other DNNs, which leads to more processing overhead. \flexisaga shows better sparse-over-dense speedups, ranging from 1.41 for ResNet50 to 4.28 for AlexNet, compared to CPU and GPU architectures using highly optimized inference runtimes. A key factor for achieving better sparse-over-dense speedups is that for each operator, the dataflow with minimal runtime was selected. Fig.~\ref{fig:flexisaga_dnn_runtimes}(b) presents the distribution of selected dataflows per DNN operator across all DNNs and all FlexiSAGA sizes. This indicates that a significant amount of sparse-over-dense speedup can be attributed to the csOS dataflow which specifically exploits the proposed CSB sparse matrix format.

\subsection{DNN Operator-wise Inference Evaluation}

\begin{figure}[!t]
	\centering
	\includegraphics[width=1.0\textwidth]{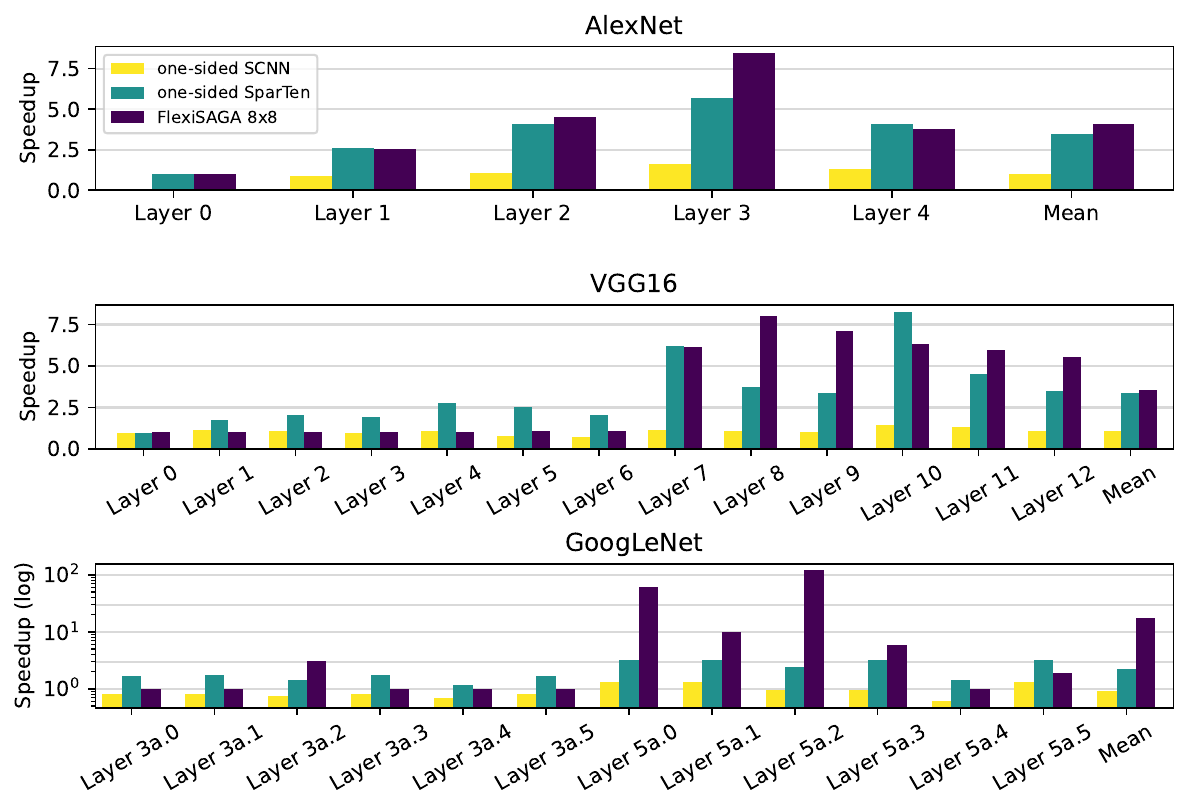}
	\caption{Operator-wise sparse-over-dense speedup comparison between one-sided SCNN, one-sided SparTen, and FlexiSAGA 8$\times$8 for AlexNet, VGG16, and GoogLeNet convolution operators.}
    \label{fig:operator_wise_sparten_scnn_flexisaga}
\end{figure}

To provide a more fine-granular analysis, this section presents an operator-wise sparse-over-dense speedup comparison of \flexisaga with the one-sided SCNN and SparTen architectures for the CONV operators of AlexNet, VGG16, and GoogLeNet (3a and 5a blocks) for which the speedups are reported in \cite{sparten2019}. Fig.~\ref{fig:operator_wise_sparten_scnn_flexisaga} shows the sparse-over-dense speedup comparison for all three architectures. Since the SCNN and SparTen architectures for which the results are reported only support CONV operators and are comprised of 64 processing elements, we compare them to \flexisaga with an SA size of 8$\times$8 and mapped only the CONV operators. While \flexisaga shows lower speedups for the first couple of CONV operators, the sparse-over-dense speedup is often higher for the second half. This can be explained by the sparsity distribution across all operators, as presented in Fig.~\ref{fig:dnn_sparsities}. This is especially apparent for the GoogLeNet operators in block 5a, where some operators have such a high sparsity that they can almost be skipped entirely, leading to a sparse-over-dense speedup of up to 100. For the mean sparse-over-dense speedup FlexiSAGA shows better results for all three DNNs.

\subsection{Design Space Exploration}

\begin{figure}[!t]
	\centering
	\includegraphics[width=1.0\textwidth]{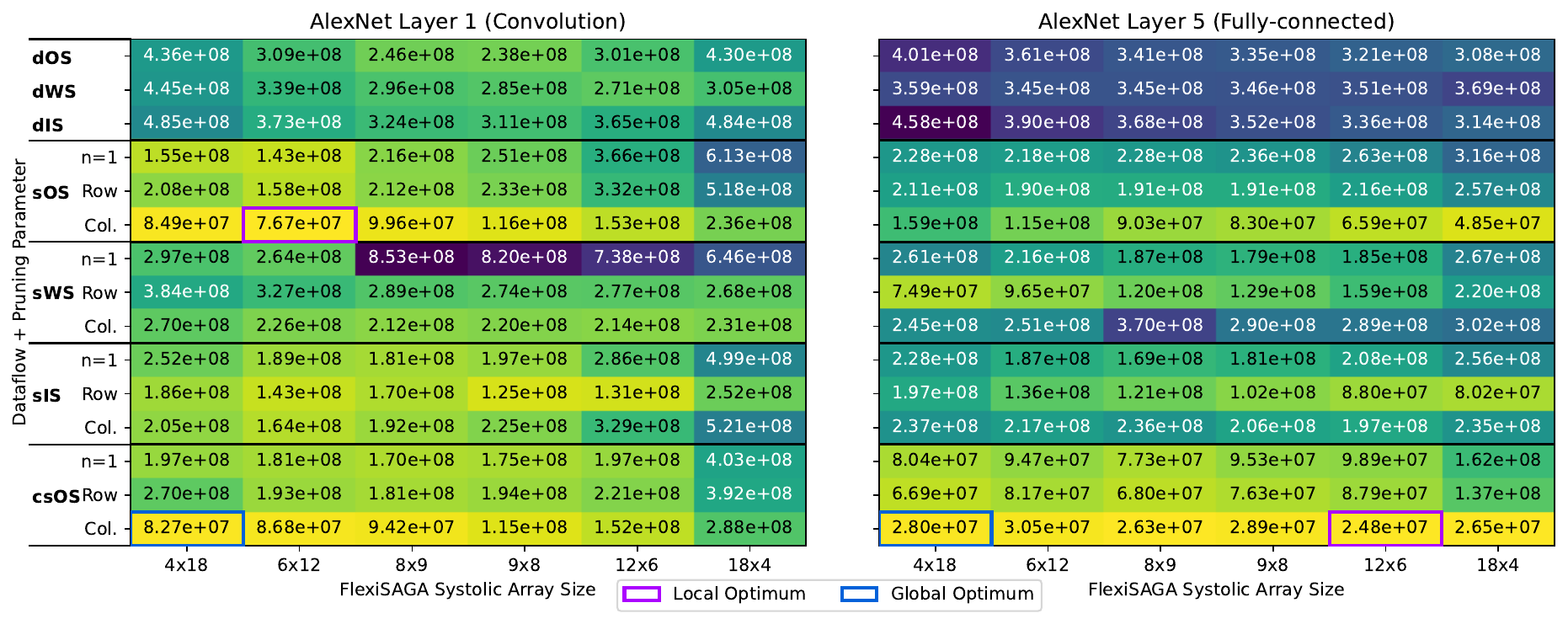}
	\caption{Design space exploration comparing the runtime in clock cycles of two AlexNet operators mapped onto a FlexiSAGA architecture of different shapes comprised of 72 PEs for all dataflows and different pruning parameters.}
    \label{fig:alexnet_dse}
\end{figure}

Lastly, we present a design space exploration (DSE) in Fig.~\ref{fig:alexnet_dse} for two AlexNet operators processed on different \flexisaga instances, all comprised of 72 PEs coupled with different dataflows and pruning parameters. The DSE heatmap shows the runtime in clock cycles. For the CONV operator on the left-hand side, the best runtime (local optimum) is achieved for a \flexisaga instance of size 6$\times$12 using the sOS dataflow with pruned column vectors of length $n=6$. For the FC operator on the right-hand side, the best runtime (local optimum) is reached with a SA of size 12$\times$6 using the csOS dataflow and pruned column vectors of length $n=12$. We conducted this DSE for all AlexNet CONV and FC operators and selected the architecture and pruning method which has the lowest whole DNN inference runtime. This resulted in a SA of size 4$\times$18 for which the DNN is pruned using column vectors with $n=4$. For the presented operators in Fig.~\ref{fig:alexnet_dse} the dataflow with the best global runtime is csOS (global optimum). The unbalanced row to column ratio of the best SA instance can be explained by the linear scaling of the memory interface. The more PEs at the border are connected to the main memory, the more data can be streamed into the SA. These results show that finding the optimal combination of architecture configuration, pruning parameters, and dataflow is not trivial and motivates the use of a DNN/HW co-design flow and an extensive performance evaluation. Generating the 1440 measurements using the VP took 421.4\,h, with a mean runtime per measurement of 17.6\,min and a standard deviation of 13.2\,min.

\section{Conclusion}
This paper presented \flexisaga a flexible systolic array GEMM accelerator for sparse and dense processing. Our results show a better sparse-over-dense whole DNN inference speedup compared to other architectures, reaching a speedup of up to 4.28. This is enabled by a near-optimal processing of DNN CONV and FC operators using seven different dense and sparse dataflows and a specifically tailored pruning technique, which constitutes a DNN/HW co-design flow.

\flexisaga is broadly applicable to DNNs comprised of mostly CONV and FC operators. Going forward, we plan to extend the range of supported operators to multi-head attention, which can be found in transformer networks such as large language and generative AI models.

\begin{credits}
\subsubsection{\ackname} This work has been funded by the German Federal Ministry of Research, Technology and Space (BMFTR) under grant numbers 16ME0129 (Scale4Edge) and 01IS22086H (MANNHEIM-FlexKI).

\end{credits}

%
%
\bibliographystyle{splncs04}
\bibliography{references}

\end{document}